\documentclass[11pt, a4paper]{article}
\usepackage{amsmath}
\usepackage{amssymb}
\usepackage[dvips]{graphics}
\usepackage[dvips]{graphicx}
\usepackage{soul}

\newcommand{\R}{\mathbb R}

\numberwithin{equation}{section}
\usepackage{color}

\def\dd{{\rm d}}
\newtheorem{theorem}{Theorem}

\newtheorem{definition}[theorem]{Definition}

\newcommand{\benumerate}{\begin{enumerate}}
\newcommand{\eenumerate}{\end{enumerate}}

\newcommand{\bitemize}{\begin{itemize}}

\newcommand{\eitemize}{\end{itemize}}

\newcommand{\der}[2]{\frac{\partial #1}{\partial #2}}

\newcommand{\dersec}[2]{\frac{\partial^{2} #1}{\partial #2^{2}}}

\newcommand{\dertot}[2]{\frac{d #1}{d #2}}

\begin{document}

\title{Thermodynamic phase transitions and shock singularities}
\author{Giuseppe De Nittis$\;^{*}$ and Antonio Moro$\;^{**}$ \\
\\
\small{$^{*}$ LAGA - Institut Galil\' {e}e - Universit\'{e} Paris 13, Villetaneuse, France} \\
\small{$^{**}$ SISSA -  International School for Advanced Studies Trieste, Italy} \\
\\
\small{{\tt email: denittis@math.univ-paris13.fr, antonio.moro@sissa.it}}
}
\date{}

\maketitle

\begin{abstract}
We show that under rather general assumptions on the form of the entropy function, the energy balance equation for a system in thermodynamic equilibrium is equivalent to 
a set of nonlinear equations of hydrodynamic type. This set of equations is integrable via the method of the characteristics and it provides the equation of state for the gas. The shock wave catastrophe set identifies the phase transition. A family of explicitly solvable models of non-hydrodynamic type such as the classical plasma and the ideal Bose gas are also discussed.

\end{abstract}

\tableofcontents 

\newpage

\section{Introduction}

A physical system in thermodynamic equilibrium is specified by a certain number of  thermodynamic variables  and  state functions. The occurrence of a  phase transition can be interpreted as the fact that there exist a thermodynamic function  possessing a certain number singular (critical) points. 

For instance, the energy balance equation for a system in thermodynamic equilibrium can be written as follows~\cite{Landau}
\begin{equation}
\label{law}
dE = T dS - P d V + \sum_{j=1}^{m} \Lambda_{j} d \tau_{j}
\end{equation}
where $E$ stays for total energy of the system, $T$ the temperature, $P$ the pressure, $S$ the entropy, $V$ the volume and $\{\tau_{j}\}_{j=1,\dots,m}$ is a set of additional parameters that determine the state of the system~\cite{Landau,Callen}. For a gas mixture of  $m$ species $\tau_{j} = N_{j}$ would be the number of particles of each species and the conjugated functions $\Lambda_{j} = \mu_{j}$ are interpreted as the associated chemical potentials. In some other cases the pairs ($\tau_{j}, \Lambda_{j}$) would play the role of a set of generalized fluxes as, for instance, mass, heat flux etc. together with their conjugated generalized potentials (see e.g.~\cite{Ruggeri, Jou}).

For sake of simplicity, let us assume that the state of the system is specified by the thermodynamic variables $V$, $T$ and $P$ only and no extra variables are involved. If from microscopic considerations one can deduce the explicit form of the Helmotz free energy $F = E - T S$  the equation of state is given by the well known formula~\cite{Landau}
\begin{equation}
\label{state0}
P + \der{F}{V}(V,T) = 0.
\end{equation}
The equation of state can be viewed as a stationary point of the Gibbs potential $\Phi = F + P V$ as a function of $V$. The constant pressure and temperature state such that the Gibbs potential has a minimum is a state of stable equilibrium. The existence of degenerate critical points of $\Phi$ such that the second and possibly higher order derivatives of $\Phi$ w.r.t. $V$ vanish detects  a phase transition. From the point of view of the singularity theory one would like to look at the Gibbs potential as a stable unfolding of the map $V  \mapsto \Phi(V,T_{c},P_{c})$  where $T_{c}$ and $P_{c}$ are suitable constant parameters at which $\Phi$ has a degenerate singularity~\cite{Yung}. The phase transition corresponds to the catastrophe set of this stable unfolding.

\begin{figure}[htbp]
\begin{center}
\includegraphics[height=4.5cm]{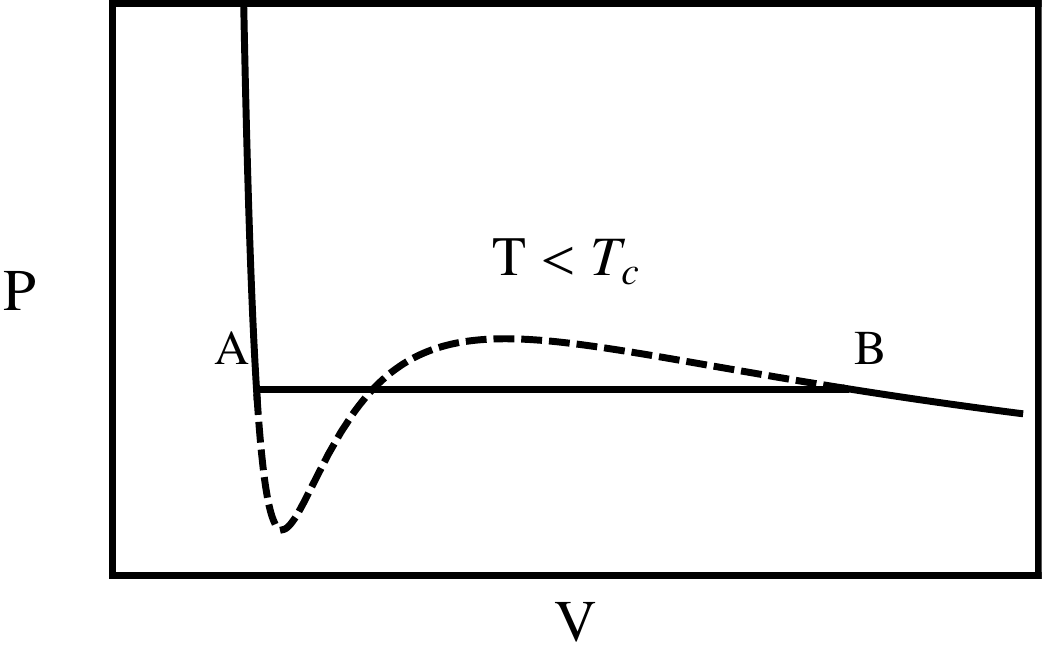}
 \end{center}
  \caption{{\footnotesize  Real gas isotherm (solid line) below the critical temperature $T_{c}$. The phase transition takes place between the points A and B where the pressure is constant. The dashed line describes a metastable state within the critical region AB as predicted by  the Van der Waals model. Outside the critical region the Van der Waals curve describes with good accuracy  several real gases.
  }}\label{intropic1}
\end{figure}

\begin{figure}[htbp]
\begin{center}
\includegraphics[height=4.5cm]{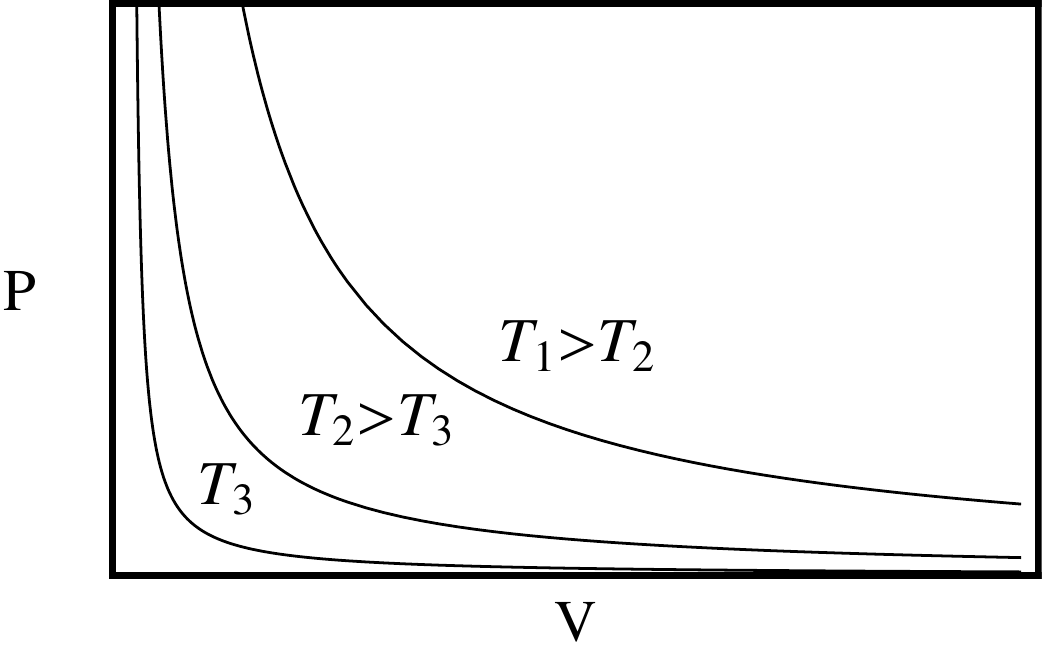}
 \end{center}
  \caption{{\footnotesize Ideal gas isotherms are given by a family of hyperbolas and there are no singular points at finite pressure and volume.
  }}\label{introidpic}
\end{figure}

A classical example of phase transition is the change of state from gas to liquid.  Figure~\ref{intropic1} shows an isothermal curve (solid line) of a real gas at a certain temperature $T$ below the critical temperature $T_{c}$.
The phase transition takes place between the points A and B where pressure remains constant as the volume increases. On the left one observes the liquid state while the gas state is observed on the right.
The ideal gas model turns out to be too simple to predict the occurrence of phase transitions as the isotherms do not possess singular points (see Figure~\ref{introidpic}). The Van der Waals model is the simplest model that preditcs a phase transitions. The dashed line in Figure~\ref{intropic1} shows the behaviour within the phase transition region according with  the Van der Waals model. It corresponds to a metastable state and, generically, it is not observed in the experiments. The correct behaviour can be recovered applying the so-called Maxwell principle~\cite{Callen}. At the phase transition the pressure remains constant and its value is determined in such a way that the area below the upper portion of the Van der Waals curve is equal to the area above the lower portion. 

\begin{figure}[htbp]
\begin{center}
\includegraphics[height=6cm]{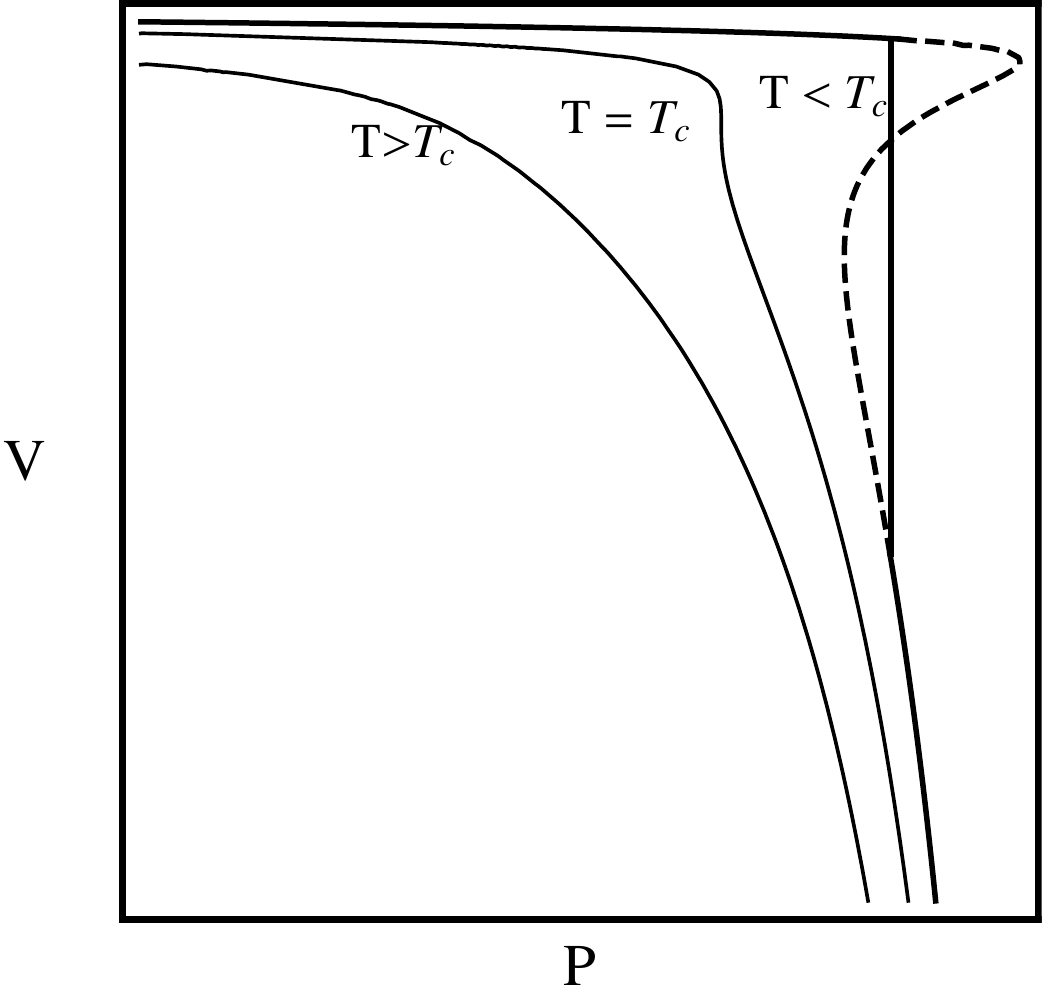}
\end{center}
  \caption{{\footnotesize Evolution at different temperatures of a Van der Waals curve (up to reflections on the plane) as a nonlinear wave solution to a hyperbolic PDE. Beyond the critical temperature $T_{c}$ the solution becomes multivalued (dashed line) and the shock takes place. The position of the jump can be recovered by the shock fitting procedure~\cite{Whitham}.
  }}\label{intropic2}
\end{figure}

This description is apparently similar to the evolution of a shock wave after the interchange of the independent variable $V$ with the dependent variable $P$ . The Figure~\ref{intropic2} shows the evolution of the Van der Waals curve (up to reflections on the plane) at different temperatures on a ($P,V$) diagram before and after the critical value $T_{c}$. In particular, at the critical temperature $T_{c}$ a gradient catastrophe occurs after which  the profile of the Van der Waals curve becomes multivalued. This analogy suggest that the  volume $V$ as a function of pressure and temperature behaves as the solution to a hyperbolic PDE of the form~\cite{Whitham}
\begin{equation}
\label{hydro0}
\der{V}{T} = \phi(V) \der{V}{P}.
\end{equation}
The solution to this equation after the critical point has a jump and consequently exists in the weak sense only. The position of the jump is prescribed by the shock fitting procedure~\cite{Whitham} in such a way the discontinuity cuts off lobes of equal area. This is nothing but the Maxwell principle mentioned above.
In the present paper we show that the function $V(T,P)$ satisfies an equation of the form~(\ref{hydro0}) under the assumption that the entropy can be decomposed into the sum of a function of $V$ plus a function of $T$. It turns out that several cases of physical interest belong to this class. Moreover, the above assumption on the functional form of the entropy can be also justified using the Boltzmann principle~\cite{Landau} for systems in the semiclassical regime.
The theory of hydrodynamic type equations of the form~(\ref{hydro0}) is well established and can efficiently be applied. In the scalar case, these equations are always integrable via the method of characteristics that provides a local implicit representation of the solution via an algebraic relation of the form
$$
\Omega(V,T,P) = 0.
$$
This relation turns out to be the equation of state of the thermodynamic system.

This approach, based on the study of a set of nonlinear hyperbolic PDEs (conservation laws) for the thermodynamic variables allows to construct the equation of state from the macroscopic features of the system such as the knowledge of a certain number of isothermal or isobaric curves, provided the functional form of the entropy fulfills certain general assumptions. This point of view shows some analogies with the one recently proposed in \cite{Cooper}.

The importance of the theory of PDEs for the study and the interpretation of thermodynamics is well established and goes back to the seminal work of C.~Carath{\'e}odory \cite{Caratheodory}. More recent
developments in this direction involve formulations in terms of contact geometry~\cite{Arnold}.
Hence, it is not surprising to come across a possible relevance in thermodynamics of the theory of integrable systems.

\medskip 

The paper is organized as follows. In Section~\ref{sec_state} we derive a set of hyperbolic PDEs of hydrodynamic type under the assumption that the entropy is separable as the sum of a function of $V$ and a function of the set of variables  ($T,\tau_{1},\dots,\tau_{m}$). The system of equations so obtained is completely integrable and its solution provides the equation of state for the system. The specific form of the entropy function and the arbitrary functions resulting from the integration can be determined on the macroscopic level in terms of a sufficient number of particular isobaric and/or isothermal curves.
In Section~\ref{sec_virial} we apply the procedure described in Section~\ref{sec_state} to the Virial expansion and discuss the Ideal and Van der Waals gas as special cases. In Section~\ref{sec_singular} we discuss the interpretation of the phase transition as the singular sector of the system of hydrodynamic type. A family of explicitly integrable cases for a class of more general separable entropy functions is discussed in Section~\ref{sec_separable}.  Section~\ref{sec_conclusions} is devoted to some concluding remarks.

\section{Equations of state}
\label{sec_state}
Let us consider the Gibbs free energy $\Phi = E - T S + V P$, then the balance equation~\eqref{law} takes the form
\begin{equation}
\label{law2}
d \Phi = - S dT + V dP +  \sum_{i=1}^{m} \Lambda_{i} d \tau_{i}
\end{equation}
where $S,V,\Lambda_1,\ldots\Lambda_m,$ are functions of the thermodynamic variables $(T,P,\tau_1,\ldots,\tau_n)\in{\cal D}\subseteq\R^{n+2}$.
The equation~(\ref{law2}) implies
\begin{equation}
\label{law_diff}
S = - \der{\Phi}{T}, \qquad\qquad V = \der{\Phi}{P}, \qquad\qquad \Lambda_{1} = \der{\Phi}{\tau_{1}} \qquad \dots \qquad \Lambda_{m} = \der{\Phi}{\tau_{m}}.
\end{equation}
The system of equations above is compatible if and only if $\Phi\in C^2({\cal D})$, i.e. the following Maxwell relations hold 
\begin{equation}
\label{comp}
\der{V}{T} = - \der{S}{P} \qquad\qquad \der{V}{\tau_{1}}=\der{\Lambda_{1}}{P} \qquad \dots \qquad \der{V}{\tau_{m}}=\der{\Lambda_{m}}{P}.
\end{equation}
We observe that, according to the second equation in~(\ref{law_diff}), the jump in the volume function $V =V(T,P)$, as in Figure~\ref{intropic2}, can be viewed as a discontinuity of the first derivative of the Gibbs free energy w.r.t. the pressure. 
\medskip

\noindent Let us assume that the entropy function $S$ and the potentials $\Lambda_i$ depend on the pressure $P$ through the volume only, that is
\begin{gather}
\begin{aligned}
\label{ansatz}
&S(T,P,\tau_1,\ldots,\tau_m)&=\ &\tilde{S}\big(V(T,P,\tau_1,\ldots,\tau_m),T,\tau_1,\ldots,\tau_m\big)\\
&\Lambda_i(T,P,\tau_1,\ldots,\tau_m)&=\ &\tilde{\Lambda}_i\big(V(T,P,\tau_1,\ldots,\tau_m),T,\tau_1,\ldots,\tau_m\big) \qquad i =1,\dots,m.
\end{aligned}
\end{gather}
Equations~(\ref{comp}) together with the assumptions~(\ref{ansatz}) give the following set of equations
\begin{gather}
\begin{aligned}
\label{comp2}
&\der{V}{T} = - \der{\tilde{S}}{V} \der{V}{P},
\qquad\qquad
&\der{V}{\tau_{i}} =  \der{\tilde{\Lambda}_{i}}{V} \der{V}{P} \qquad i =1,\dots, m.
\end{aligned}
\end{gather}
Introducing the notation
\begin{align*}
&\phi_{0}(V,T,\tau_{1},\ldots,\tau_m) = - \left(\der{\tilde{S}}{V} \right)^{-1}\\ 
&\phi_{i}(V,T,\tau_{1},\ldots,\tau_m) = \phi_{0}(V,T,\tau_{1},\ldots,\tau_m)\ \der{\tilde{\Lambda}_{i}}{V}  \qquad i =1,\dots,m
\end{align*}
equations~(\ref{comp2}) look like as follows
\begin{gather}
\label{comp3}
\begin{aligned}
&\der{V}{P}\ =\ \phi_{0}(V,T,\tau_{1},\ldots,\tau_m)\ \der{V}{T} \\ \vspace{1mm}
&\der{V}{\tau_{i}}\ =\ \phi_{i}(V,T,\tau_{1},\ldots,\tau_m)\ \der{V}{T} \qquad i =1,\dots,m.
\end{aligned}
\end{gather}
The system of equations~(\ref{comp3}) is compatible, i.e. $\partial_{\tau_{i}} \partial_{P} V = \partial_{P} \partial_{\tau_{i}} V$, and it is integrable via the method of characteristics. The general solution is a scalar field $V(T,P,\tau_{1},\dots,\tau_{m})$ constant along the family of characteristic curves on the space of variables $(T,P,\tau_{1},\dots,\tau_{m})$. Characteristics are defined via the system
\begin{gather}
\label{characteristics}
\begin{aligned}
&\der{T}{s} = 1 \\
&\der{P}{s} = \frac{1}{\phi_{0}(V(P(s), T(s), \tau_{1}(s),\dots,\tau_{m}(s)) )} \\
&\der{\tau_{i}}{s} = \frac{1}{\phi_{i}(V(P(s), T(s), \tau_{1}(s),\dots,\tau_{m}(s)) )}, \qquad i =1, \dots,m
\end{aligned}
\end{gather}
where $s$ parametrizes the curves.
\medskip

\noindent Under the extra assumption that the dependence of the entropy function is separable with respect to the volume $V$ and the set of variables  $(T, \tau_{1},\ldots,\tau_m)$ 
as follows
\begin{equation*}
\tilde{S}(V,T,\tau_{1},\ldots,\tau_m) = \varphi_{0}(V) + \psi_{0}(T,\tau_{1},\ldots,\tau_m),\qquad \Lambda_{i}(V,T,\tau_{1},\ldots,\tau_m) =  \varphi_{i}(V) + \psi_{i}(T,\tau_{1},\ldots,\tau_m)
\end{equation*}
the set of equations~(\ref{comp3}) takes the so-called hydrodynamic  form
\begin{gather}
\label{comp4}
\begin{aligned}
&\der{V}{P}\ =\ \phi_{0}(V)\ \der{V}{T} \\ \vspace{1mm}
&\der{V}{\tau_{1}}\ =\ \phi_{1}(V)\ \der{V}{T}\\
&\qquad\qquad\vdots\\
&\der{V}{\tau_{m}}\ =\ \phi_{m}(V)\ \der{V}{T}.
\end{aligned}
\end{gather}
The characteristic speeds $\phi_{0},\dots, \phi_{m}$ depend in fact on the variable $V$ only.
The general solution to the system~(\ref{comp4}) is locally given by the formula
\begin{equation}
\label{state}
T + \phi_{0}(V) P + \sum_{i =1}^{m} \phi_{i}(V) \tau_{i} = f(V)
\end{equation}
where $f(V)$ is an arbitrary function of its argument (for a review on the general theory of hydrodynamic type systems see e.g.~\cite{Dubrovin} and references therein). We refer to  the equation~(\ref{state}) as the \emph{equation of state}. If the characteristic speeds $\phi_{1}, \dots, \phi_{m}$, are known, the function $f(V)$ can be specified either by fixing a particular isothermal curve (or isotherm), $T =T_{0}, \tau_{i} =\tau_{0i}$
or an isobaric curve (or isobar) $P =P_{0}, \tau_{i} =\tau_{0,i}$ in the space of thermodynamic variables $(V, P,T, \tau_{1},\dots,\tau_{m})$.
We note that assigning the volume on a particular isobaric curve is equivalent to assigning an initial value problem for the system~(\ref{comp4}). Alternatively, fixing the isothermal curve means to solve the initial value problem for the same system~(\ref{comp4}) written in evolutionary form with respect to the set of times~$T,\tau_{1},\dots,\tau_{m}$.
For instance, choosing the initial datum $V_0(P)=V(T_0,P,\tau_{0,1},\ldots,\tau_{0,m})$, on the isothermal curve  
 $T =T_{0}, \tau_{1} = \tau_{0,1},\dots,\tau_{m} = \tau_{0,m}$, we have, at least locally, 
$$
f(V) = T_{0} + \phi_{0}(V) P_{0}(V)  + \sum_{i=1}^{m} \phi_{i}(V)\ \tau_{0,i},
$$
where $P_{0} = V_{0}^{-1}$ is the inversion of the initial datum $V_0(P)$. The relation above specifies the form of the function $f(V)$ in terms of the initial datum.
We also note that the \emph{isochoric} hypersurfaces $V = const$ on the $n+2$ dimensional space ${\cal D}$
are, by definition, the characteristics of the system~(\ref{comp4}). We also note that if, apart from $f(V)$, a certain number of characteristic speeds is not known, these can be obtained by measuring the same number of additional isotherms and/or isobars.
For sake of simplicity assume that all the functions $\phi_{i}$, $i=0,1, \dots,m$ have to be determined. Given $m+2$ distinct isotherms in the form $V_i(P)=V(T_i,P,\tau_{i1},\ldots,\tau_{im})$, one can locally compute the set of inverse functions $P_{0} = V_{0}^{-1}, \dots, P_{m} = V_{m}^{-1}$. Then, the functions $f(V)$ and $\phi_{i}(V)$, $i=0,1,\dots,n$ are given by the solution to the following linear system
$$
\begin{aligned}
&T_{0} + \phi_{0}(V) P_{0}(V) + \sum_{i=1}^{m} \phi_{i}(V) \tau_{0i} = f(V) \\
&T_{1} + \phi_{0}(V) P_{1}(V) + \sum_{i=1}^{m} \phi_{i}(V) \tau_{1i} = f(V) \\
& \dots \\
&T_{m+1} + \phi_{0}(V) P_{m}(V) + \sum_{i=1}^{m} \phi_{i}(V) \tau_{mi} = f(V).
\end{aligned}
$$

\medskip

For the particular choice $\phi_{j}(V) = c_{j} V^{j+1}$ with $j = 0,\dots, m$, where $c_{j}$'s are arbitrary constants, the system~(\ref{state}) coincides with the first $m+1$ members of the \emph{Burgers-Hopf hierarchy}~\cite{Zakharov}
\begin{gather}
\label{comp5}
\begin{aligned}
&\der{V}{P} = c_{0}\ V\ \der{V}{T} \\ \vspace{1mm}
&\der{V}{\tau_{1}} = c_{1}\ V^2\ \der{V}{T}\\
&\ldots\\
&\der{V}{\tau_{n}} = c_{m}\ V^{m+1}\ \der{V}{T}.
\end{aligned}
\end{gather}
In this case the equation of state is 
\begin{equation}
\label{state_hopf}
T + c_{0} V P + \sum_{j =1}^{m} c_{j} V^{j+1} \tau_{j} = f(V).
\end{equation}
The whole hierarchy is simply obtained by taking $j =1,\dots, \infty$.


\section{Virial expansion}
\label{sec_virial}
In the present section we discuss the class of equations of state in the form of the virial expansion. For sake of simplicity we discuss first the ideal gas example and then the Van der Waals model as a particular case of the more general virial expansion in the case no extra variables $\tau_{j}$ are involved. Applying the procedure outlined above we show how, given either two isobars or equivalently two isotherms, one can recover the form of the characteristic speeds and  determine uniquely the equation of state as a particular solution to the  equation of hydrodynamic type.

\subsubsection*{Ideal gas}
Let us consider a monoatomic gas of constant number of particles $N$ and equation of state
\begin{equation*}
T + \phi_{0}(V) P = f(V).
\end{equation*}
Assume we are given two particular isobaric curves $V_{1}(T) =V(T, P_{1})$ and $V_{2}(T) =V(T, P_{2})$ with $P_{1} \neq P_{2}$ of the form
\begin{equation}
\label{perf1}
V_{1}(T) = \frac{T}{c_{1}},\qquad\qquad V_{2}(T) = \frac{T}{c_{2}},
\end{equation}
where $c_{1}$ and $c_{2}$ are  two given constants.
The equation of state along these isobars can be written as
\begin{equation*}
T_{j}(V) + \phi_{0}(V) P_{j} = f(V), \qquad j =1,2
\end{equation*}
where $T_{j} = V_{j}^{-1}$ are obtained by inverting \eqref{perf1}.
Using the equations~(\ref{perf1}) we get the following system of linear equations for the functions
$\phi_{0}(V)$ and $f(V)$
\begin{gather}
\label{ideal_isobars}
\begin{aligned}
&c_{1} V + \phi_{0}(V) P_{1} = f(V) \\
&c_{2} V + \phi_{0}(V) P_{2} = f(V).
\end{aligned}
\end{gather}
Subtracting the equations~(\ref{ideal_isobars}) we get
\begin{equation}
\label{perf1b}
\phi_{0} (V) = - \frac{V}{\alpha}
\end{equation}
where 
$$
\alpha = \frac{P_{1} - P_{2}}{c_{1} - c_{2}}
$$ 
is a constant that has to be independent on the particular pair of chosen isobars. 
Equation \eqref{perf1b} implies that the entropy is a logarithmic function of the volume $V$. We will show, at the end of this section, how this property can be  heuristically understood in terms of microscopic considerations. Moreover, one can see that the constant 
$\alpha$ should be proportional to the  total number of particles $N$, in such a way that  
$\alpha = n R$ where $n = N/N_{A}$ is  the molar number, $N_{A}$ the Avogradro's number and $R$ the universal gas constant.
Using the relation~(\ref{perf1b}) in~(\ref{ideal_isobars}) one obtains
\begin{equation}
\label{ideal_fam}
f(V) = s V, \qquad\qquad \textup{where} \qquad\qquad  s = \frac{P_{1} c_{2} - P_{2} c_{1}}{P_{1} - P_{2}}.
\end{equation}
The constants $\alpha$ and $s$ specify the physical properties of the gas.
The physical constraint
$$
\lim_{V \to \infty} P(T,V) = 0
$$
implies that $s = 0$ or equivalently
$$
\frac{P_{1}}{P_{2}} =\frac{c_{1}}{c_{2}}.
$$
This fixes uniquely the solution and provides equation of state for the {\it ideal gas}
\begin{equation}
\label{ideal_BH}
P V = n R T.
\end{equation}
One can easily check that proceeding as showed in the previous section, the same result can be obtained by fixing two isotherms of the form
\begin{equation}
\label{ideal_iso}
P_{1}(V) = \frac{c_{1}}{V} \qquad\qquad P_{2}(V) = \frac{c_{2}}{V}.
\end{equation}

\subsubsection*{\bf Real gas} 
Let us assume that two particular isotherms on the $(V,P)$ plane have the following form
\begin{equation}
\label{virial_iso}
P_{j}(V) = \frac{1}{V -\nu} \left(\alpha_{0}(T_{j}) + \frac{\alpha_{1}(T_{j})}{V} + \dots + \frac{\alpha_{k}(T_{j})}{V^{k}}  \right), \qquad j = 1,2
\end{equation}
for an arbitrary positive integer $k$. In the relation~(\ref{virial_iso}) it is required that $0\leqslant\nu <V$ where $\nu$  is interpreted as the \emph{intrinsic volume}  occupied by the molecules of the gas . We observe that the physical conditions
$$
\lim_{V \to \infty} P_{j}(V) = 0 \qquad  \qquad \lim_{V\to0} P_{j}(V) = \infty, \qquad \qquad j = 1,2
$$
are satisfied. The particular choice
$$
\alpha_{0} = n R T \qquad \qquad \nu = 0 \qquad \qquad \alpha_{1}(T) =\dots = \alpha_{k}(T) = 0
$$
gives the ideal gas isotherms~(\ref{ideal_iso}).
Proceeding as in the previous example, solving the system of linear equations
\begin{equation*}
T_{j} + \phi_{0}(V) P_{j}(V) = f(V), \qquad j =1,2
\end{equation*}
together with the assumption that $\phi_{0}(V)$ is a function of the volume only, we have that the virial coefficients $\alpha_{i}(T_{j})$ can be at most linear functions, i.e.
$$
\alpha_{i}(T_{j}) = \gamma_{i} T_{j} + c_{i}, \qquad i =0,1,\dots,k
$$
and
\begin{align}
&\phi_{0}(V) = - (V - \nu) \left( \gamma_{0} + \frac{\gamma_{1}}{V} + \dots + \frac{\gamma_{k}}{V^{k}} \right)^{-1} \\
& f(V)  = - \frac{c_{0} V^{k} + c_{1} V^{k-1} + \dots + c_{k}}{\gamma_{0} V^{k} + \gamma_{1} V^{k-1} + \dots + \gamma_{k}}.
\end{align}
The equation of state reads as
\begin{equation}
\label{virial_state}
T - (V - \nu) \left(\gamma_{0} + \frac{\gamma_{1}}{V}  + \dots + \frac{\gamma_{k}}{V^{k}} \right)^{-1} + \frac{c_{0} V^{k} + c_{1} V^{k-1} + \dots + c_{k}}{\gamma_{0} V^{k} + \gamma_{1} V^{k-1} + \dots + \gamma_{k}} = 0.
\end{equation}
{\bf Van der Waals approximation.} Virial expansion  of the order $k=2$ with
\begin{align*}
&\nu = n b& \qquad &\gamma_{0} = n R& \qquad &\gamma_{1} =\gamma_{2} = 0& \\
&c_0=0 &\qquad& c_{1} = - n^{2} a& \qquad &c_{2} = n^{3} b
\end{align*}
provides the Van der Waals equation of state
\begin{equation*}
\left(P + \frac{n^{2}  a}{V^{2}} \right) (V - n b) = n R T.
\end{equation*}
This is a particular implicit solution to the equation
\begin{equation}
\label{VdW_PDE}
\der{V}{P}  + \frac{V - n b}{n R} \der{V}{T} = 0
\end{equation}
that is equivalent to the Burgers-Hopf equation up to a Galilean change of variables.

\subsubsection*{\bf Remark 1: Entropy function of a semiclassical gas} 
In the examples discussed above, the assumption of separability of the entropy function w.r.t. the variables $T$ and $V$ together with the two particular isobars (or isotherms) of the form~(\ref{virial_iso}) implies that the entropy has to be logarithmic in the volume $V$. This results can be heuristically understood taking into account the definition of the entropy function for a semiclassical gas (Boltzmann principle)~\cite{Landau} 
\begin{equation}\label{ent2}
S = \kappa \log(\Delta\Gamma)= \kappa \log\left(\frac{\Delta q\ \Delta p}{(2\pi\hslash)^{N r}}\right)=\log\left(\Delta q\right)+\log\left(\Delta p\right) - N\ r \log\left(2\pi\hslash\right)
\end{equation}
where $\Delta\Gamma=\Delta q\ \Delta p/(2\pi\hslash)^{N r}$ is the statistical weight, $N$ the number of particles and $r$ the number of degrees of freedom of a particle.
$\Delta \Gamma$ counts the number of distinct microscopic states within the allowed region $\Delta q\ \Delta p$ of the phase space. 

 If the gas is assumed to be confined in a macroscopic volume $V$  and occupies an intrinsic volume  $\nu$ then  $\Delta q=(V-\nu)^N$. 
Morever, as the temperature is a measure of the mean kinetic energy of the system, we can assume in first instance that
$$
\log\left(\Delta p\right) = \psi_0(T)
$$ 
where $\psi_{0}$ is a certain function. Hence, we end up with the following form of the entropy function
\begin{equation}\label{ent1}
S(T,P)=\tilde{S}\big(V(T,P),T\big)=N \kappa \log\big(V(T,P)-\nu\big)+ \psi_0(T)+ c
\end{equation}
where $\kappa$ is identified with the Boltzmann constant $\kappa = R /N_{A}$ and $N_{A}$ is Avogadro's number. 
Finally, one can reasonably assume that $\nu\propto N$ as in the Van der Waals approximation. In particular one can set $\nu=(N/N_A)b$ where $b$ measures the physical volume occupied by a gas mole. As the ideal gas is made of point-like particles it is $b=0$.


\section{Singular sector and phase transitions}
\label{sec_singular}
Let us introduce the function
\begin{equation}
\label{Omega}
\Omega(V,T,P, \tau_{1},\dots,\tau_{m}) =T + \phi_{0}(V) P + \sum_{i =1}^{m} \phi_{i}(V) \tau_{i} - f(V).
\end{equation}
Hence, the equation
$$
\Omega(V,T,P, \tau_{1},\dots,\tau_{m}) = 0
$$
gives locally the solution $V(T,P,\tau_{1},\dots,\tau_{m})$ to the system~(\ref{comp4}) provided the invertibility condition $\partial \Omega/\partial V \neq0$ is satisfied. In the following we will adopt the notation
\begin{align*}
&\Omega'  = \der{\Omega}{V} \qquad \Omega'' = \dersec{\Omega}{V} \qquad \Omega''' = \frac{\partial^{3}\Omega}{\partial V^{3}} \qquad \Omega^{(l)} = \frac{\partial^l\Omega}{\partial V^l},\qquad\qquad l = 4,5,6,\dots.
\end{align*}
Following~\cite{Kod_Kon}, let us introduce the singular sector of codimension $s \leq n+2$ for the system of equations~(\ref{comp4}) 
\begin{equation}
\label{sector}
Z_{s}  = \left \{ (T,P,\tau_{1},\dots,\tau_{m}) \in \mathbb{R}^{m+2} \; \; | \; \; V \in {\cal U}_{j} \;\; \textup{such that} \;\; s = \sum_{j=1}^{m+2} j \; | {\cal U}_{j} | \right \}
\end{equation}	
where 
\begin{equation*}
{\cal U}_{j} = \left\{ V >0 \; | \; \Omega^{(l)}= 0 \quad \forall\   0 \leq l \leq j\qquad  \text{and}\qquad \Omega^{(j+1)} \neq 0 \right\}.
\end{equation*}
and $|{\cal U}_{j}|$ is the number of solutions considered in ${\cal U}_{j}$. In the following it will be useful the

{\definition[Critical sector]  The \textup{critical sector} is the singular sector of highest codimension. The critical sector is said to be \textup{maximal} if its codimension is $m+2$.}

\medskip

\noindent The critical sector identifies the occurrence of the phase transition as well as the shock formation of the solution to the system~(\ref{comp4}).

\subsubsection*{Van der Waals case}
For a Van der Waals gas we have $\phi_{1} = \phi_{2} = \dots = \phi_{m} \equiv 0$ and the system of equations~(\ref{comp4}) reduces to the equation~(\ref{VdW_PDE}).  In this case we have
\begin{equation}
\label{wdW-ss1}
\Omega(V,T,P) =\frac{\tilde{\Omega}(V,T,P)}{n RV^2}=  T - \frac{V - n b}{n R} \; P  - \frac{n a}{R} \left(\frac{1}{V} - \frac{n b}{V^{2}} \right)
\end{equation}
where $\tilde{\Omega}$ is a polynomial function of $V$ defined as follows
$$
\tilde{\Omega}(V,T,P): =  - P V^{3} + (n b P + n R T) V^{2} - n^{2} a V + n^{3} a b. 
$$
The singular sector $Z_{1}$  is specified by the conditions
$$
\tilde{\Omega} = 0  \qquad \qquad \qquad \tilde{\Omega}' = 0 \qquad \qquad \qquad \tilde{\Omega}'' \neq 0.
$$
These conditions give the set 
\begin{equation}
\label{VdW_cod1}
Z_{1} = Z_{1}^{+}  \cup Z_{1}^{-},\qquad \text{with}\qquad Z_{1}^{\pm} = \left \{ (T,P) \; | \; \tilde{\Omega}(V^{\pm}(T,P), T, P ) = 0 \right\}
\end{equation}
where $V^{\pm}(T,P)$ are real roots of the quadratic polynomial $\tilde{\Omega}'$, i.e.
$$
V^{\pm}(T,P) =\frac{-2 n b P - 2 n R T \pm \sqrt{-12 n^{2} a P + (2 n b P+ 2 n R T)^{2}}}{6 P}.
$$
\begin{figure}[htbp]
\begin{center}
\fbox{
\includegraphics[height=7cm]{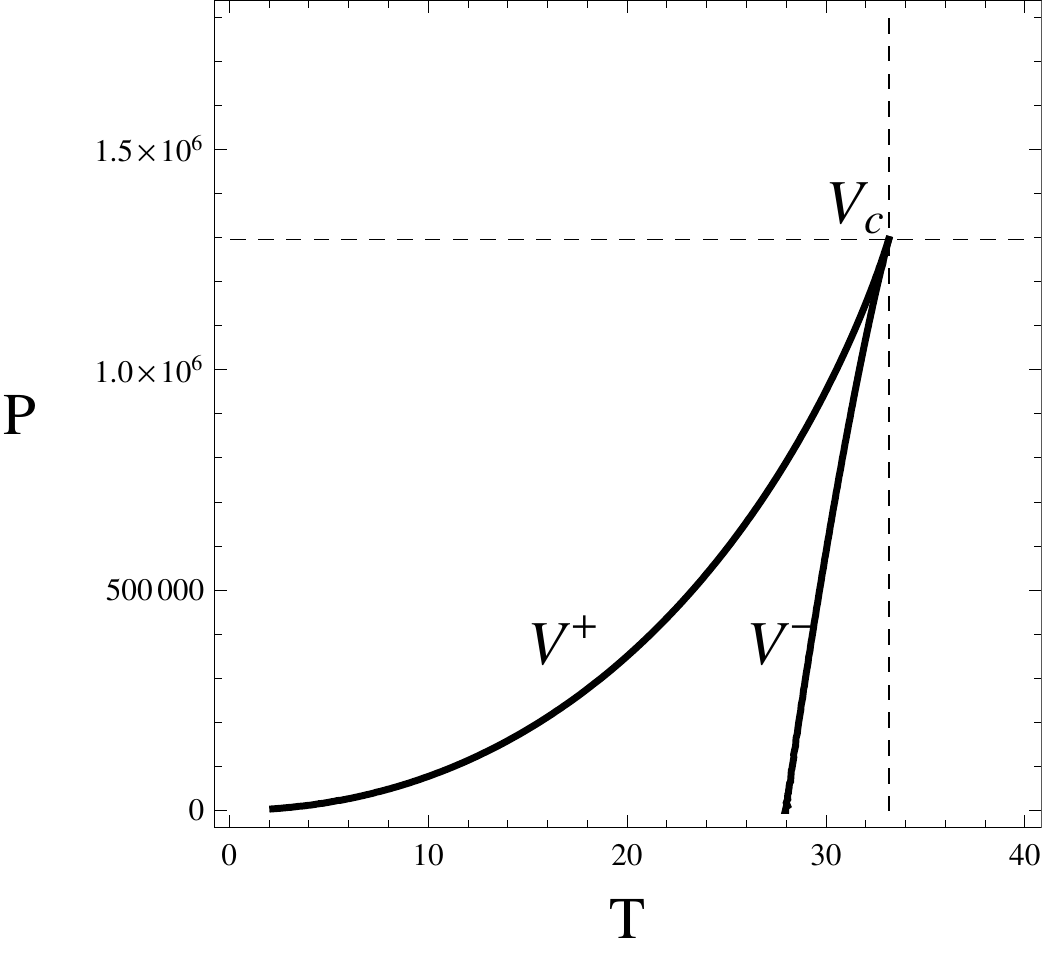}
}
\end{center}
  \caption{{\footnotesize Singular sector $Z_{1}$ for the hydrogen gas according with the Van der Waals model: $a =2.476 \times 10^{-2} \; m^{6} Pasc/mol^{2}$, $b = 2.661 \times 10^{-5} \; m^{3}/mol$, $n = 10^{3}$, $R = 8.3144 \; J/K mol$. The intersection of the two lines is  the critical sector $Z_{2}$$=\{T_{c} =33.159 \; K $, $P_{c} =1.29508 \times 10^{6} \; Pasc \}$.
  }}\label{VdW_Z1}
\end{figure}
As shown in figure~(\ref{VdW_Z1}) the singular sector $Z_{1}$ is a collection of two curves emerging from the {\it critical point} 
$$
Z_{1}^{+}  \cap Z_{1}^{-}=\left\{(T_{c},P_{c})\right\}=\left\{
\left (\frac{8 a}{27 b R}, \; \frac{a}{27 b^{2}} \right)
\right\}.
$$
\medskip
\noindent The singular sector $Z_{2}$ is given by the set of equations
$$
\tilde{\Omega}= 0 \qquad \qquad \qquad  \tilde{\Omega}' = 0 \qquad \qquad \qquad  \tilde{\Omega}''= 0.
$$
Note that $\tilde{\Omega}^{(3)}  = - 6 P \neq 0$.
More explicitly one has
\begin{equation}
\label{VdW_cod2}
Z_{2}  = \left \{ (T_c,P_c) \right\}.
\end{equation}
Hence, $Z_{2}$ is the critical sector  of a first order phase transition for the Van der Waals gas. The corresponding critical value of the volume is 
$$
V_c = 3 n b.
$$
In this case the critical sector  $Z_{2}$ is maximal. As shown in Figure~\ref{VdW_pic} the critical isotherm  has an inflection point due to the fact that the first non vanishing derivative is odd or, equivalently, the codimension of the critical sector is even.
\begin{figure}[htbp]
\begin{center}
\fbox{
\includegraphics[height=7cm]{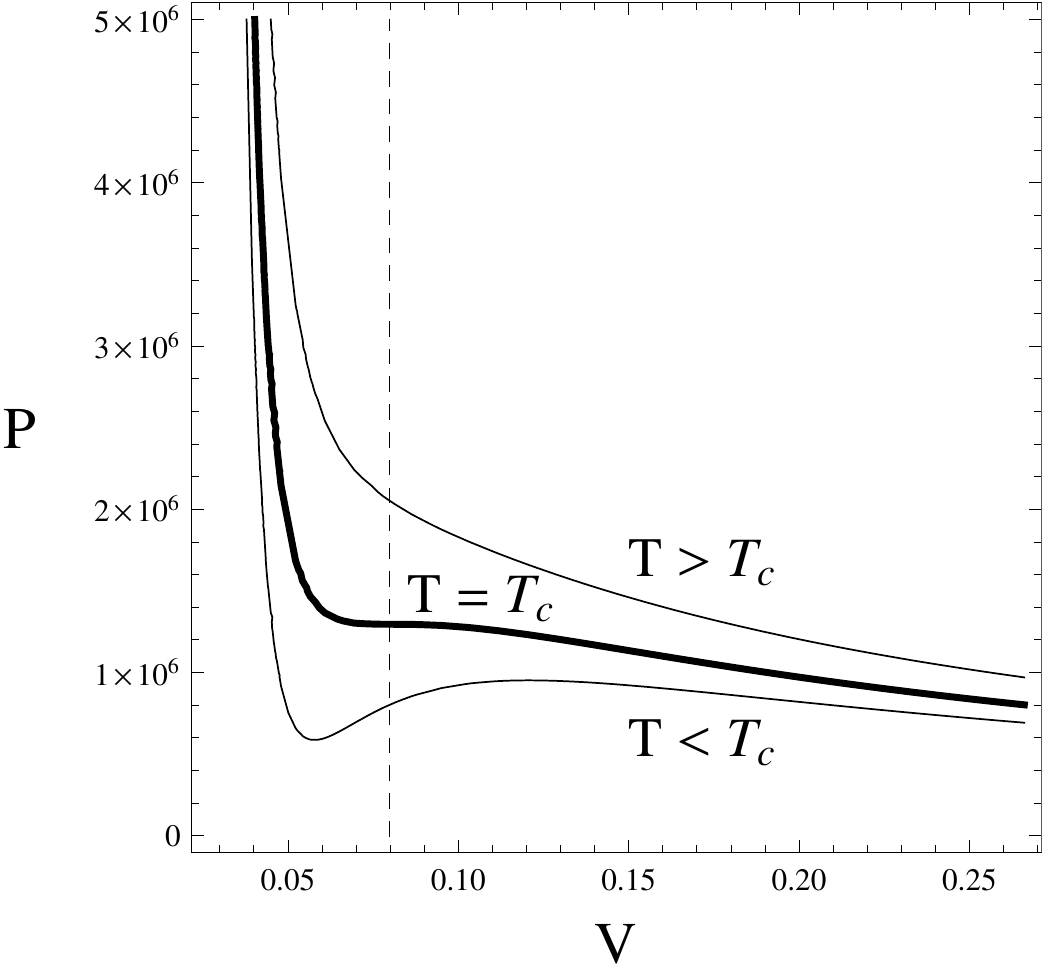}
}
\end{center}
  \caption{{\footnotesize The bold line shows the critical isotherm for the hydrogen gas according with the Van der Waals model: $a =2.476 \times 10^{-2} \; m^{6} Pasc/mol^{2}$, $b = 2.661 \times 10^{-5} \; m^{3}/mol$, $n = 10^{3}$, $R = 8.3144 \; J/K mol$. The critical point is ($V_{c} =0.07983 \; m^{3}$, $T_{c} =33.159 \; K $, $P_{c} =1.29508 \times 10^{6} \; Pasc$). Below the critical isotherm there are two stationary points where the isotherm has a local maximum and a local minimum. On the ($T,P$) plane stationary points  describe the singular sector $Z_{1}$  in Figure~\ref{VdW_Z1}.
  }}\label{VdW_pic}
\end{figure}

\medskip

\subsection*{Deformed Virial Expansion}
Let us consider the function
\begin{equation}
\label{van_def1}
\Omega(V,T,P,\tau_{1},\dots,\tau_{m}) =T - \frac{V - n b}{n R} P + \sum_{j=1}^{m} \phi_{j}(V) \tau_{j} -f(V).
\end{equation}
The state of the system is specified by the set of variables $(T,P,V,\tau_{1},\dots,\tau_{m})$.  Let us define
$$
\mu = \max \{k \in \mathbb{N} \; \;  | \; \;  {\cal U}_{k}  \neq \emptyset \}.
$$
Note that $\mu \leq m+2$ as for $k > m+2$ the system of equations 
$$
\Omega = \Omega' =\dots = \Omega^{(k)} = 0.
$$
is overdetermined. For $V \in {\cal U}_{\mu}$ the system of equations
\begin{gather}
\label{Umu}
\begin{aligned}
&\Omega = T - \frac{V - n b}{n R} P + \sum_{j=1}^{m} \phi_{j}(V) \tau_{j} - f(V)  = 0\\
&\Omega' = - \frac{P}{n R} + \sum_{j=1}^{m} \phi_{j}'(V) \tau_{j} - f'(V) = 0 \\
&\Omega'' =  \sum_{j=1}^{m} \phi_{j}''(V) \tau_{j} - f''(V) = 0 \\
&\dots \\
&\Omega^{(\mu)} =  \sum_{j=1}^{m} \phi_{j}^{(\mu)}(V) \tau_{j} - f^{(\mu)}(V) = 0.
\end{aligned}
\end{gather}
can be viewed as a linear system in the set of $m+2$ variables ($T,P,\tau_{1},\dots,\tau_{m}$).

If $\mu < m+ 2$, solving the system~(\ref{Umu}) with respect to the set of variables ($T,P,\tau_{1},\dots,\tau_{\mu-1}$) one can parametrize the critical sector $Z_{\mu}$ as follows
\begin{align*}
&T = T(V,\tau_{\mu}, \dots, \tau_{m}) \\
&P = P(V,\tau_{\mu}, \dots, \tau_{m}) \\
&\tau_{1} = \tau_{1} (V,\tau_{\mu}, \dots, \tau_{m}) \\
&\dots \\
&\tau_{\mu-1} = \tau_{\mu-1} (V,\tau_{\mu}, \dots, \tau_{m}).
\end{align*}
In particular, for $\mu = m+1$, $Z_{\mu}$ is a curve in the $m+2$ dimensional affine space ($T,P,\tau_{1},\dots,\tau_{m}$).

\begin{figure}[htbp]
\begin{center}
\fbox{
\includegraphics[height=5cm]{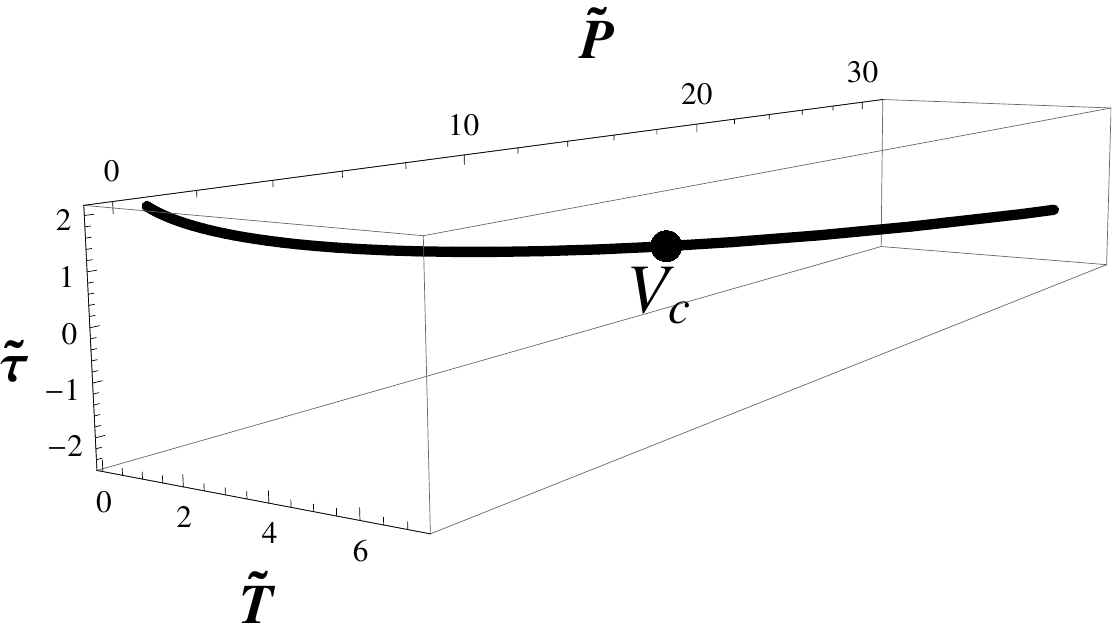}
}
\end{center}
  \caption{{\footnotesize Critical sector for the deformed Van der Waals model in the case $m =1$ with $a =2.476 \times 10^{-2} \; m^{6} Pasc/mol^{2}$, $b = 2.661 \times 10^{-5} \; m^{3}/mol$, $n = 10^{3}$, $R = 8.3144 \; J/K mol$. Rescaled variables $\tilde{T} = T \times 10^{-1}$, $\tilde{P} = P \times 10^{-5}$ and $\tilde{\tau} = \tau_{1}$ are used. The critical point is ($T_{c} =33.159 \; K $, $P_{c} =1.29508 \times 10^{6} \; Pasc$, $\tau_{c} = 0 \; K \; Pasc /J \; m^{3}$) and corresponds to the critical volume $V_{c} =0.07983 \; m^{3}$.
  }}\label{defVdW_pic}
\end{figure}

If $\mu = m+2$ the critical sector is maximal. It is given by a set of isolated points
$$
T = T(V_{k}) \qquad P = P(V_{k}) \qquad \tau_{1} = \tau_{1} (V_{k}) \qquad \dots \qquad \tau_{m} = \tau_{m}(V_{k})
$$
where $V_{k}$ is the $k$-th root of
\begin{align*}
 \Omega^{(m+2)} \left (V, T(V), P(V),\tau_{1}(V),\dots, \tau_{m}(V) \right) = 0.
\end{align*}
As a particular example let us consider the following deformed Virial expansion
\begin{equation}
\label{defVirial}
f(V) = \sum_{j=1}^{m} \gamma_{j} \phi_{j}(V) \qquad \phi_{j}(V) = V^{-j} \qquad 
\end{equation}
where $\gamma_{j}$ are arbitrary real constants. The rank on the system~(\ref{Umu}) is maximal and the system is compatible. For the particular choice
$$
\gamma_{1}  = \frac{n a}{R} \qquad \qquad \gamma_{2} = - \frac{n^{2} a b}{R} \qquad \qquad \gamma_{j} = 0 \;\;\; j =3,\dots,m
$$
one obtains a deformed Van der Waals model. Let us analyze the one parameter deformation $m=1$.
In this case $\mu = 2$, the conditions $\Omega =$ $\Omega' =$ $\Omega'' = 0$ give
\begin{equation}
\label{defVdW_Z2}
T = \frac{n^{2} a b}{R V^{5}} (3 V - n b) \qquad \qquad P = \frac{n^{3} a b}{V^{3}} \qquad \qquad \tau_{1} = \frac{n a}{R V} (V - 3 n b)
\end{equation}
and
$$
\Omega''' (V, T(V),P(V),\tau_{1}(V)) = - \frac{6 n^{2} a b }{R V^{5}} \neq 0.
$$
The critical sector  $Z_{2}$ is the curve~(\ref{defVdW_Z2}) in the parameter space ($T,P,\tau_{1}$). The figure~(\ref{defVdW_pic}) shows the critical sector for the hydrogen gas. The plain circle on the curve corresponds to the critical volume $V_{c} = 3 n b$ for the Van der Waals model such the $\tau_{1} = 0$.

\subsection*{Remark 2: Unfoldings}

Locally, in the vicinity of the critical point, the Van der Waals equation is associated with the universal unfolding of codimension 2 of a Riemann-Hugoniot catastrophe~\cite{Yung}. More precisely, the universal unfolding gives the Gibbs free energy
$$
\Phi = F + PV
$$
where $F = E - TS$. The deformed virial expansion discussed above can be interpreted as a higher codimension unfolding of the Van der Waals model.


\section{Separable entropy functions}
\label{sec_separable}
Let us consider a system the state of which is completely specified by the set of thermodynamic variables $(T,P,V)$. Assume we are given a more general separable entropy function of the form
\begin{equation}
S(T,P)=\tilde{S}\big(V(T,P),T \big) = \sum_{j =0}^{\infty} \sigma_{j} \big(V(T,P)\big) \frac{ \partial \rho_{j} }{\partial T} (T)
\end{equation}
where $\sigma_{j}(V)$ and $\rho_{j}(T)$ are arbitrary functions of their argument. 
The system~(\ref{comp3}) reduces to the following equation
\begin{equation}
\label{nonhydro}
\der{V}{T}(T,P) + \sum_{j=0}^{\infty} \frac{\partial \sigma_{j}}{\partial V}\big(V(T,P) \big)\ \frac{\partial \rho_{j}}{\partial T}(T)\  \der{V}{P}(T,P) = 0.
\end{equation}
 The characteristic curves $P(T)$ are defined by the equation
 \begin{equation}
 \label{nonhydro_char}
 \dertot{P}{T}(T) = \sum_{j=0}^{\infty} \frac{\partial \sigma_{j}}{\partial V}\big(V\big(T,P(T)\big) \big)\ \frac{\partial \rho_{j}}{\partial T}(T).
 \end{equation}
 As the solution $V$ to the equation~(\ref{nonhydro}) is constant along the characteristics, integration of~(\ref{nonhydro_char}) leads to the following implicit formula of the solution
 \begin{equation}
 \label{nonhydro_state}
 P - \sum_{j=0}^{\infty} \frac{\partial \sigma_{j}}{\partial V}\big(V\big(T,P\big) \big)\ \; \rho_{j} (T)  = f\big(V\big(T,P\big) \big)
 \end{equation}
 where $f(V)$ is an arbitrary function of its argument. As discussed above the function $f(V)$ can be fixed by specifying either a particular isobare or a particular isotherm.
 
 \medskip

 \section*{Examples}
Few cases of physical interest that can be obtained as a virial expansion with coefficients depending on the temperature fit in the class of separable entropy functions discussed above. We consider here some explicit examples.

 \medskip

\noindent{\bf Classical Plasma.}
The case of completely ionized gas (\emph{classical plasma}) can be recovered  setting  $\sigma_{j} = \rho_{j} = 0$ for $j \geq 2$, $f(V) = 0$ and
$$
\sigma_{0}(V) = nR\ \log V
\qquad \rho_{0}(T) = T \qquad 
\sigma_{1}(V) =  \frac{2c}{\sqrt{V}} \qquad  \rho_{1}(T) = \frac{1}{\sqrt{T}} 
$$
wher $c$ is a certain constant. The
 equation~(\ref{nonhydro_state}) gives the equation of state for the classical plasma \cite{Landau}
\begin{equation}
P= \frac{n RT}{V} -\frac{c}{V^{3/2}\sqrt{T}}.
\end{equation}

\medskip
 
\noindent{\bf Redlich--Kwong equation.} 
Setting  $\sigma_{j} = \rho_{j} = 0$ for $j \geq 2$, $f(V) = 0$ and
$$
 \sigma_{0}(V) = n R\log(V-nb) \qquad \rho_{0}(T) = T \qquad \sigma_{1}(V) = \frac{nc}{b}\log\left(\frac{V+nb}{V}\right) \qquad  \rho_{1}(T) =  \frac{1}{\sqrt{T}}
$$
where $c$ and $b$ are some constants, the equation~(\ref{nonhydro_state}) gives
 the  Redlich--Kwong equation of state \cite{Redlich}
\begin{equation}\label{eq:redich}
P=\frac{nRT}{V-nb}-\frac{n^2c}{\sqrt{T}\ V(V+nb)}
\end{equation}
Equation \eqref{eq:redich} usually provides a more accurate description of real gas phase transition than the Van der Waals equation.

\medskip
 
\noindent{\bf Peng--Robinson equation.}
Setting  $\sigma_{j} = \rho_{j} = 0$ for $j \geq 2$, $f(V) = 0$ and
\begin{align*}
 &\sigma_{0}(V) = n R\log(V-nb) && \rho_{0}(T) = T &&\\
 &\sigma_{1}(V) = -\frac{a}{\sqrt{2} n b} \textup{arctanh} \left [ \frac{V + n b}{\sqrt{2} n b} \right]&& \rho_{1}(T) =  \big[1+c(T_c-T)\big]^2
\end{align*}
where $a,c$ and $T_c$ are constants, the equation~(\ref{nonhydro_state}) gives the equation
\begin{equation}
P=\frac{nRT}{V-nb}-\frac{a\big[1+c(T_c-T)\big]^2}{V^2+2nb V-(nb)^2}
\end{equation}
that is known as Peng--Robinson equation of state \cite{Peng}. This equation  arise in the theory of  nonpolar liquids.

\medskip
 
\noindent{\bf Dieterici equation.}
 The choice
$$
 \sigma_{j}(V) =n R\int \frac{\dd V}{(V-nb)V^j} \qquad \rho_{j}(T) = \frac{1}{j!}\left(-\frac{a}{R}\right)^jT^{1-j}
  \qquad f(V) = 0
$$
with $j=0,1,2,\ldots$ and $a$ a suitable constant provides the Dieterici equation of state~\cite{Dieterici}
\begin{equation}
P=\frac{nRT}{(V-nb)}\ {\rm e}^{-\frac{a}{RTV}}.
\end{equation}
A straightforward computation shows that
$$
 \sigma_{j}(V) =  \frac{n R}{(nb)^j}\log\left(\frac{V-nb}{V}\right) + \sum_{r=1}^{j-1}\frac{n R}{r(nb)^{j-r}V^r}\qquad \text{if}\qquad j> 1.
$$

\medskip
 
\noindent{\bf Ideal Bose gas equation.} Setting  $\sigma_{j} = \rho_{j} = 0$ for $j \geq 1$, $f(V) = 0$ and
$$
\sigma_0(V)=nR\log(V)\qquad \rho_0(T)=\frac{T_{c} \textrm{Li}_{\alpha+1}\big(z(T)\big)}{\zeta(\alpha)} \left(\frac{T}{T_c}\right)^{\alpha+1}
$$
where $z(T)= {\rm e}^{\frac{c}{T}}$, $\alpha$ and $c$ are constants depending on the physical properties of the system $\textrm{Li}_{\alpha+1}$ is the function polylogarithm, $\zeta$ is the Riemann zeta function, and $T_c$ is the critical temperature at which a Bose--Einstein condensate begins to form.
Equation~(\ref{nonhydro_state}) results into the ideal Bose gas \cite{Landau} state equation
\begin{equation}
    P=\frac{nRT}{V}~\frac{\textrm{Li}_{\alpha+1} \left (z(T) \right)}{\zeta(\alpha)} \left(\frac{T}{T_c}\right)^\alpha.
\end{equation}

\section{Concluding remarks}
\label{sec_conclusions}
We have shown that a large class of equations of state for real gases can be obtained as solutions of a set of integrable hyperbolic PDEs. This set of equations is equivalent to the balance equation~(\ref{law}) under the assumption that the entropy function is separable in the variables $V$ and $T$. The simplest case leads to a set of equations of hydrodynamic type for which the characteristic speeds are function of the dependent variable $V$ only. The specific functional form of the unknown characteristic speeds as well as the arbitrary functions resulting from the integration procedure can be explicitly determined using a suitable number of particular isothermal or isobaric curves. 

For more general separable entropy functions the hyperbolic conservation law can still be explicitly integrated. This case contains few examples of equation of state for real gases that can be viewed as a virial expansion with coefficients depending on the temperature.

The equations of state predict the occurrence of a phase transition that is understood as the shock wave dynamics of solutions to  a hyperbolic PDE. The critical point where the phase transition starts to develop corresponds  to the gradient catastrophe set where the shock start to form. The singularity and catastrophe theory~\cite{Arnold2,Yung} turns out to be the natural language for the study of local properties of a system near the critical point. Further analysis of the deformed virial expansion involves the study of stable unfoldings of higher codimension.

The approach discussed above can be generalized for the study of composite systems in thermal contact and multiphase thermodynamic systems to be associated with  a multicomponent system of hydrodynamic PDEs.

A more accurate description of the thermodynamic functions near the phase transition should concern the study of more general entropy functions depending on first and possibly higher derivatives of V. This will produce diffusive and/or dispersive higher order equations.
In fact, it should be emphasized that  discontinuous solutions, such as in Figure~\ref{intropic2}, can be obtained as the small diffusion limit of the Burgers equation \cite{Whitham, MCL}.
This suggest that in proximity of the phase transition higher order effects should play a role in the  description of the real phase transition.

\section*{Acknowledgements}
We are pleased to thank Boris Dubrovin, Boris Konopelchenko, Kenneth McLaughlin, Tamara Grava, Andrea Raimondo for useful discussions and comments. A. M. is supported by ERC Grant From-PDE (P.I. Boris Dubrovin) and the Grant ``Giovani Ricercatori della SISSA" (P.I.  A.M.). G.D. is supported by the grant ANR-08-BLAN-0261-01. A.M. and G.D. are partially supported by the Grant of Istituto Nazionale di Alta Matematica - GNFM, Progetti Giovani 2011 (P.I. A.M.).

\end{document}